# PERFORMANCE COMPARISON OF LINK, NODE AND ZONE DISJOINT MULTI-PATH ROUTING STRATEGIES AND MINIMUM HOP SINGLE PATH ROUTING FOR MOBILE AD HOC NETWORKS

Natarajan Meghanathan

Jackson State University, 1400 Lynch St, Jackson, MS, USA
natarajan.meghanathan@jsums.edu

## ABSTRACT

*The high-level contribution of this paper is a simulation-based analysis to evaluate the tradeoffs between lifetime and hop count of link-disjoint, node-disjoint and zone-disjoint multi-path routes vis-à-vis single-path minimum hop routes for mobile ad hoc networks. The link-disjoint, node-disjoint and zone-disjoint algorithms proposed in this paper can be used to arrive at benchmarks for the time between successive multi-path route discoveries, the number of disjoint paths per multi-path set and the hop count per multi-path set. We assume a multi-path set exists as long as at least one path in the set exists. Simulation results indicate that the number of zone-disjoint paths per multi-path set can be at most 2, which is far lower than the number of node and link-disjoint paths available per multi-path set. Also, the time between zone-disjoint multi-path discoveries would be far lower than the time between node and link-disjoint multi-path route discoveries and can be at most 45% more than the time between single minimum-hop path route discoveries. However, there is no appreciable difference in the average hop counts per zone-disjoint, node-disjoint and link-disjoint multi-path sets and it can be only at most 15% more than the average minimum hop count determined using single-path routing. We also observe that even though the number of link-disjoint paths per multi-path set can be as large as 35-78% more than the number of node-disjoint paths per multi-path set, the time between two successive link-disjoint multi-path discoveries can be at most 15-25% more than the time between two successive node-disjoint multi-path discoveries, without any significant difference in the hop count per multi-path set.*

## 1. INTRODUCTION

A mobile ad hoc network (MANET) is a dynamic distributed system characterized by node mobility, limited battery power of nodes and limited channel bandwidth. Due to the limited transmission range of the nodes, MANET routes are often multi-hop in nature and a node assists its peers in route discovery and data propagation. MANET routing protocols are of two types: reactive and proactive. Reactive or on-demand routing protocols (e.g., Dynamic Source Routing – DSR [1], Ad hoc On-demand Distance Vector Routing –AODV [2]) use a network-wide flooding of route request messages to build and maintain routes, but only when needed. Proactive routing protocols (e.g., Destination Sequenced Distance Vector routing – DSDV [3]) tend to maintain routes between all pairs of nodes all the time and hence in the presence of a dynamically changing topology, incur considerable route maintenance overhead compared to on-demand protocols [4]. Hence, most of the recent research in MANETs is on reactive on-demand routing and we restrict ourselves to this routing technique in this paper.

On-demand routing protocols incur high route discovery latency and also incur frequent route discoveries in the presence of a dynamically changing topology. Recent research has started to focus on multi-path routing protocols for fault tolerance and load balancing. Multi-path on-demand routing protocols tend to compute multiple paths, at both the traffic sources as well as at intermediary nodes, in a single route discovery attempt. This reduces both the route discovery latency and the control overheads as a route discovery is needed only when all the discovered paths fail. Spreading the traffic along several routes could alleviate congestion and bottlenecks. Multi-path routing also provides a higher aggregate bandwidth and effective load balancing as the data forwarding load can be distributed over all the paths.

Multi-paths can be of three types: link-disjoint, node-disjoint and zone disjoint. For a given source *s* and destination *d*, the set of link-disjoint *s-d* routes comprises of paths that have no link present in more than one constituent *s-d* path. Similarly, the set of node-disjoint *s-d* routes comprises of paths that have no node (other than the source and destination) present in more than one constituent *s-d* path. A set of zone-disjoint *s-d* routes comprises of paths such that an intermediate node in one path is not a neighbour node of an intermediate node in another path. Multi-path on-demand routing protocols tend to compute multiple paths between a source-destination (*s-d*) pair, in a single route discovery attempt. A new network-wide route discovery operation is initiated only when all the *s-d* paths fail. To the best of our knowledge, there is no detailed simulation study on the stability and average hop count of link-disjoint, node-disjoint and zone-disjoint multi-path routes vis-à-vis the minimum hop single path routes in a centralized fashion. Our work establishes benchmarks for the time between successive multi-path route discoveries for each of the three routing strategies, the average hop count per path used from a multi-path set and the number of paths per multi-path set.

For a given source *s* and destination *d*, the multi-path set of link (or node, or zone) disjoint *s-d* routes at a given time instant is determined as follows: Determine the minimum-hop *s-d* path on the current network graph and add it to the set of link (or node, or zone) disjoint routes. Remove the links (or the intermediate nodes, or the intermediate nodes as well as their neighbours except the source and destination) that constituted the just determined *s-d* path from the network graph and repeat the above procedure until no more *s-d* routes are available. We assume the *s-d* routes in a multi-path set are used in the increasing order of the hop count. In other words, the *s-d* route with the least hop count is used as long as it exists, then the *s-d* route with the next highest hop count is used as long as it exists and so on. We thus persist with the determined multi-path set of *s-d* routes as long as at least one path in the set exists. We also determine the sequence of minimum-hop single path *s-d* routes over the duration of a network simulation session and use it as a benchmark to observe the tradeoff between the stability and average hop count of multi-path routes.

The rest of the paper is organized as follows: In Section 2, we discuss related work in the area of multi-path routing in MANETs and review the protocols proposed for link, node and zone-disjoint routing. Section 3 introduces the algorithms we use to determine the set of link-disjoint, node-disjoint and zone-disjoint routes for the duration of a network simulation session. In Section 4, we describe our simulation environment and present simulation results comparing the performance of link-disjoint, node-disjoint and zone-disjoint multi-path routes vis-à-vis minimum hop single path routes. Section 5 concludes the paper. Throughout the paper, we use the terms ‘path’ and ‘route’ interchangeably. They mean the same.

## 2. LITERATURE REVIEW ON MANET MULTI-PATH ROUTING

In a typical on-demand single path routing protocol like DSR or AODV, the source node, when it does not have the route to send data to a destination node, initiates a route discovery process using flooding. The source node broadcasts a route-request (RREQ) message, tagged with a sequence number, in its neighbourhood. An intermediate node receiving a RREQ message (that originated from a given source node and a sequence number) will broadcast the message in its neighbourhood exactly once. The RREQ messages will propagate along different routes to the destination. The destination will pick up the RREQ message that propagated along a route that best satisfies the route selection metrics of the routing protocol and send a unicast route reply (RREP) along the selected route back to the source.

Multi-path routing protocols proposed for ad hoc networks make use of the propagation of the RREQ messages along several paths to the destination and let the destination to send RREP along more than one path. The routing protocols avoid the RREP storm by selecting only few of the different paths. Since nodes communicate through the shared wireless medium, the selected paths need to be as independent as possible in order to avoid transmissions from a node along one path interfering with transmissions on a different path. The aggregate bandwidth achieved with multi-path routing may not be the sum of the bandwidth of the individual paths. Metrics such as correlation and coupling factor are used to calculate the relative degree of independence among the multiple paths [5]. The correlation factor, measured only for node-disjoint paths, indicates the number of links connecting two node-disjoint paths. The coupling factor, measured for both node-disjoint and link-disjoint paths, is defined as the average number of nodes that are blocked from receiving data on one of the paths when a node in the other path is transmitting. Node-disjoint routes offer the highest degree of fault tolerance and aggregate bandwidth. Network topology and channel characteristics (measured through the correlation and coupling factors) have been observed to severely limit the gain obtained from multi-path routing [6].

In [7], the authors advocate the need to consider similarity among the multiple *s-d* paths with that of the shortest *s-d* path and stress the need to use similar paths for multi-path data propagation. Routing using multiple paths similar to the shortest path will reduce the chances of out-of-order packet delivery and also result in lower end-to-end delay per packet. The authors in [8] develop an analytical model for evaluating the effectiveness of multi-path routing. They show that unless we use a very large number of paths, the load distribution with multi-path routing is almost the same as in single path routing. An efficient approach for generalized load balancing in multi-path packet switched networks was proposed in [29]. In [30], we had studied the impact of different MANET mobility models on link and node disjoint multi-path routing.

The three multi-path routing strategies can be ranked as follows, in the increasing order of independence: link-disjoint routing, node-disjoint routing and zone-disjoint routing. It may not be always possible to simultaneously send data across two link-disjoint paths or two node-disjoint paths as the transmission of data in a link that is part of one path may require a node that is part of another path to remain idle (controlled by the channel access mechanism). It has been observed earlier [9] that larger the correlation factor between two node-disjoint paths, the larger will be the average end-to-end delay for both the paths and also the larger will be the difference in the end-to-end delay along the two paths. If two link-disjoint or node-disjoint routes are physically close enough to interfere with each other during data communication, the nodes in these multi-path routes may constantly contend for accessing the shared channel and the multi-path routing protocol may end up performing worse than any single path routing protocol [10]. In [11], the authors argue that benefits (improvement in throughput and reduction in end-to-end delay) obtained with multi-path routing become insignificant with respect to single path routing if we take into consideration the interference between the multiple paths and the cost of discovering these paths. Thus, multi-path routing may not be a sound strategy if the constituent multiple paths suffer interference among themselves. This motivates the need to consider zone-disjoint path routing also as a potentially effective multi-path routing strategy because the intermediate nodes of the zone-disjoint paths are not located in the neighbourhood of each other and zone-disjoint paths have a coupling factor of zero.

Zone-disjoint routing with directional antennas has been observed to yield a significant improvement in network throughput and reduction in end-to-end delay compared to zone-disjoint routing using omni-directional antennas [10]. Zone-disjoint routing has a non-zero correlation factor in an omni-directional network system. With omni-directional antennas, a source node will not be able to simultaneously transmit data to more than one of its neighbours and a destination node will not be able to simultaneously receive data from more than one of its neighbours. Hence, zone-disjoint paths exhibit a correlation factor of 2 in an omni-directional antenna system. Nevertheless, this value is far less than the correlation factors observed for link-disjoint and node-disjoint paths [12]. The correlation factor of zone-disjoint paths in a directional antenna system is zero as each node could set its transmission to only the target node. The zone-disjoint paths in a directional antenna system are thus 100% independent as one can simultaneously send data on all of these paths.

### 2.1 Review of Link-Disjoint Multi-path Routing Protocols

Multi-path routing protocols for MANETs are mostly either multi-path extensions of DSR or AODV. In Split multi-path routing (SMR) [13], the intermediate nodes forward RREQs that are received along a different link and with a hop count not larger than the first received RREQ. The destination selects the route on which it received the first RREQ packet (which will be a shortest delay path), and then waits to receive more RREQs. The destination node then selects the path which is maximally disjoint from the shortest delay path. If more than one maximally disjoint path exists, the tie is broken by choosing the path with the shortest hop count.

The Ad hoc On-demand Multipath Distance Vector (AOMDV) routing protocol [14] is an extension of AODV to compute multiple loop-free link-disjoint routes. The RREQs that arrive via different neighbours of the source node define the maximum number of node-disjoint/link-disjoint paths that are possible. For every destination node *d*, an intermediate node *i* maintains the list of next hop nodes, the hop count for the different paths to the destination node *d* and the "advertised hop count"(the maximum hop count for all paths from *i* to *d*), with respect to the latest known sequence number for *d*. An intermediate node accepts and forwards a route advertisement as an alternate path to the destination only if the route advertisement came from a neighbour node that has not yet sent the route advertisement for the destination sequence number and the hop count in the route advertisement is less than the advertised hop count to the destination. When a node receives a route advertisement for the destination with a greater sequence number, the next hop list and the advertised hop count values are reinitialized. The destination node replies for the RREQs arriving from unique neighbours. A multi-path routing scheme that extends AOMDV by using a traffic-path allocation scheme has been proposed in [15] and it is based on cross-layer measurements of path statistics that reflects the queue size and congestion level of each path. The scheme utilizes the Fast Forward (FF) MAC forwarding mechanism [16] to reduce the effects of self-contention among frames at the MAC layer.

### 2.2 Review of Node-Disjoint Multi-path Routing Protocols

The AODV-Multi-path (AODVM) routing protocol [17] is an extension of the AODV protocol to determine node-disjoint routes. An intermediate node does not discard duplicate RREQ packets and records them in a RREQ table. The destination responds with an RREP for each RREQ packet received. An intermediate node on receiving the RREP, checks its RREQ table and forwards the packet to the neighbour that lies on the shortest path to the source. The neighbour entry is then removed from the RREQ table. Also, whenever a node hears a neighbour node forwarding the RREP packet, the node removes the entry for the neighbour node in its RREQ table.

More recently, a geographic multi-path routing protocol (GMP) [18] has been proposed to reduce interference due to route coupling. The RREQ will have information regarding the locations of the first hop and the last hop intermediate nodes on the path. The destination chooses the path through which it first received the RREQ. For a subsequently received RREQ, the destination measures the distance between the first hops of the path traversed by this RREQ and the already selected paths and also the distance between the last hops of the path traversed by this RREQ and the already selected paths. If both these distances are greater than twice the transmission range of the nodes, the path traversed by the received RREQ is selected.

EMRP is an energy-aware multi-path routing protocol [19] that considers the available energy and the forwarding load at the intermediate nodes of the multiple paths before distributing the load across them. The destination node replies with a RREP packet for each RREQ packet. An intermediate node receiving the RREP packet updates information regarding the distance between the node and the next hop node, the number of retransmission attempts corresponding to the last successful transmission, the current queue length, the current remaining energy of the node. The source node then computes a weight for each route through which the RREP traversed. Routes with minimum weight are preferred as such routes have more remaining energy, less energy consumption due to transmission and reception, less crowded channel in the neighbourhood of the nodes in the path and more bandwidth available.

### 2.3 Review of Zone-Disjoint Multi-path Routing Protocols

The Zone-Disjoint Multi-path extension of the Dynamic Source Routing (ZD-MPDSR) protocol [20] proposed for an omni-directional system works as follows: Whenever a source node has no route to send data to a destination node, the source node initiates broadcast of the Route-Request (RREQ) messages. The number of active neighbours for a node indicates the number of neighbour nodes that have received and forwarded the Route Request (RREQ) message during a route discovery process. The RREQ message has an *ActiveNeighbourCount* field and it is updated by each intermediate node before broadcasting the message in the neighbourhood. When an intermediate node receives the RREQ message, it broadcasts a 1-hop RREQ-query message in its neighbourhood to determine the number of neighbours who have also seen the RREQ message. The number of RREQ-query-replies received from the nodes in the neighbourhood is the value of the *ActiveNeighbourCount* field updated by a node in the RREQ message. The destination node receives several RREQ messages and selects the node-disjoint paths with lower *ActiveNeighbourCount* values and sends the Route-Reply (RREP) messages to the source along these paths. Even though the selection of the zone-disjoint paths with lower number of active neighbours will lead to reduction in the end-to-end delay per data packet, the route acquisition phase will incur a significantly longer delay as RREQ-query messages are broadcast at every hop (in addition to the regular RREQ message) and the intermediate nodes have to wait to receive the RREQ-query-reply messages from their neighbours. This will significantly increase the control overhead in the network.

In order to reduce the route acquisition delay associated with ZD-MPDSR, a Cluster-based Zone Multi-path Dynamic Source Routing (CZM-DSR) protocol was proposed in [21]. Here, an intermediate node upon receiving a RREQ message records the number of times it has seen the message in a locally maintained *ActiveNeighbourCount* variable in memory and broadcasts the message further if it has been seen for the first time. The destination node sends back a Route-Reply (RREP) message to the source node for every RREQ received. The path traced by the RREQ message is included in the RREP message. When an intermediate node receives the RREP message, it includes its *ActiveNeighbourCount* value in the message and forwards the message to the next hop node on the path towards the source. The source receives RREP messages through several paths and chooses the path whose maximum value for the *ActiveNeighbourCount* is the minimum. However, CZM-DSR will still incur a larger control message overhead and possibly a RREP-storm as the destination node would send a RREP message for every RREQ message received.

A Multi-path Distance Vector Zone Routing Protocol (MDVZRP) for MANETs has been proposed in [22]. When a new node (say node *i*) joins a network, it broadcasts a Hello beacon message to its immediate neighbours. The neighbour nodes on receiving the Hello message update their routing table with a new entry for the sender of the message, node *i*, and in turn send their entire routing table (full dump) to their new neighbour, node *i*. In addition, the one-hop neighbour nodes broadcast the Hello message to their own neighbours (i.e., to the 2-hop neighbours of node *i*). This process of broadcasting the Hello message is repeated by every node if the node falls within the zone radius centred at node *i*. MDVZRP uses the notion of zone radius (measured in terms of the number of hops) to restrict the scope of the broadcast of the Hello message. Every *k*-hop neighbour node ($1 \leq k \leq$ zone radius) that receives the Hello message updates its routing table by adding an entry for the originating node of the Hello message and the neighbour node from which the Hello message was received the first time is included as the next hop node. Meanwhile, using the routing tables received from all of its neighbour nodes, the new node determines a set of node disjoint paths to every node in the entire network. If a distant node, say node *j*, falls outside the zone radius of a node, say node *i*, to which *j* wants to send data packets, then node *j* initiates a RREQ-based broadcast route search. When an intermediate node receives the RREQ message and it has a valid route (i.e., the next hop node information) for the targeted destination node of the RREQ message, the intermediate node sends back a RREP to the originating source node of the RREQ message. Such intermediate nodes are located at the periphery of the proactive routing zone centred at the targeted destination node. The RREQ message is thus not propagated all the way to the destination node. The source node, upon receiving RREPs from several of the peripheral nodes, learns the set of node-disjoint routes to the destination and starts sending the data packets through these routes.

A Cluster-Based Multi-Path Routing (CBMPR) protocol has been recently proposed in [23]. Nodes are organized into clusters – the radius of each cluster is two to three hops. Each cluster is controlled by a clusterhead that is responsible for gathering the link state information from all its member nodes, constructing the cluster topology and advertising the cluster topology information back to the member nodes. Intra-cluster communication is managed through link-state routing, while inter-cluster communication is through gateway nodes that are present in both the clusters. When a source node in one cluster has to determine multiple disjoint paths to a destination node in another cluster, it sends a RREQ message to its clusterhead, which further broadcasts the message to the clusterheads of its adjacent clusters. The RREQ message propagation is continued all the way to the cluster in which the destination node is located. The destination node receives RREQ messages across several paths whose constituent nodes are the clusterheads. The destination node selects the clusterhead-disjoint paths (i.e., disjoint-paths in which a clusterhead does not appear more than once) and sends back the RREP messages to the source through these paths. The throughput obtained with CBMPR has been observed to increase proportionally with respect to the number of clusterhead-disjoint paths used, illustrative of the independence between these paths.

While determining a maximally zone-disjoint multi-path between a source-destination (*s-d*) pair, it is imperative to consider all the active routes (between every *s-d* pair) in the system rather than only considering the zone-disjoint paths between the particular source *s* and destination *d*. In [12], the authors have proposed a trial and error algorithm to determine two maximally zone-disjoint shortest paths between an *s-d* pair. The algorithm is based on determining an initial set of node-disjoint paths between the *s-d* pair and then iteratively discarding the *s-d* path that has the largest value for the hop count * correlation factor with all of the other active routes in the system.

A 3-directional zone-disjoint multi-path routing protocol (3DMRP) has been proposed in [24]. 3DMRP discovers up to three zone-disjoint paths (one primary path and two secondary paths) based on a greedy forwarding technique at a reduced control overhead without using directional antennas. The two secondary paths, if exist, are discovered by avoiding the RREP overhearing zone created during the acquisition of the primary path.

# 3. ALGORITHMS TO DETERMINE THE SET OF LINK-DISJOINT, NODE-DISJOINT AND ZONE-DISJOINT MULTI-PATHS

We now explain the algorithms to determine the sequence of link-disjoint, node-disjoint and zone-disjoint paths for MANETs. Let $G$ ($V$, $E$) be the graph representing a snapshot of the network topology collected at the time instant in which we require a set of link-disjoint, node-disjoint or zone-disjoint routes from a source node $s$ to a destination node $d$. Note that $V$ is the set of vertices (nodes) and $E$ is the set of edges (links) in the network. We say there is a link between two nodes if the distance between the two nodes is less than or equal to the transmission range of the nodes. We assume all nodes are homogeneous and have identical transmission range.

Figures 1, 2 and 3 respectively illustrate the algorithms to determine the set of link-disjoint, node-disjoint and zone-disjoint *s-d* routes on a graph $G$ collected at a particular time instant. Let $P_L$, $P_N$ and $P_Z$ be the set of link-disjoint, node-disjoint and zone-disjoint *s-d* routes respectively. We use the Dijkstra $O(n^2)$ algorithm to determine the minimum hop *s-d* path in a graph of $n$ nodes. If there exist at least one *s-d* path in $G$, we include the minimum hop *s-d* path $p$ in the sets $P_L$, $P_N$ and $P_Z$.

To determine the set $P_L$ (refer Figure 1), we remove all the links that were part of $p$ from the graph $G$ to obtain a modified graph $G^L$ ($V$, $E^L$). We then determine the minimum hop *s-d* path in the modified graph $G'$, add it to the set $P_L$ and remove the links that were part of this path to get a new updated $G^L$ ($V$, $E^L$). We repeat this procedure until there exists no more *s-d* paths in the network. The set $P_L$ is now said to have the link-disjoint *s-d* paths in the original network graph $G$ at the given time instant.

**Input:** Graph $G$ ($V$, $E$), source $s$ and destination $d$
**Output:** Set of link-disjoint paths $P_L$
**Auxiliary Variables:** Graph $G^L$ ($V$, $E^L$)
**Initialization:** $G^L (V, E^L) \leftarrow G (V, E)$, $P_L \leftarrow \varphi$.
**Begin**
    **While** ($\exists$ at least one *s-d* path in $G^L$)
        $p \leftarrow$ Minimum hop *s-d* path in $G^L$.
        $P_L \leftarrow P_L \cup \{p\}$
        $\forall_{edge, e \in p}$ $G^L (V, E^L) \leftarrow G^L (V, E^L - \{e\})$
    **end While**
    **return** $P_L$
**End**

**Figure 1:** Algorithm to Determine the Set of Link-Disjoint *s-d* Paths in a Network Graph

To determine the set $P_N$ (refer Figure 2), we remove all the intermediate nodes (nodes other than the source $s$ and destination $d$) that were part of the minimum hop *s-d* path $p$ in the original graph $G$ to obtain the modified graph be $G^N$ ($V^N$, $E^N$). We determine the minimum hop *s-d* path in the modified graph $G^N$ ($V^N$, $E^N$), add it to the set $P_N$ and remove the intermediate nodes that were part of this *s-d* path to get a new updated $G^N$ ($V^N$, $E^N$). We then repeat this procedure until there exists no more *s-d* paths in the network. The set $P_N$ is now said to contain the node-disjoint *s-d* paths in the original network graph *G*.

---

**Input:** Graph $G(V, E)$, source *s* and destination *d*
**Output:** Set of node-disjoint paths $P_N$
**Auxiliary Variables:** Graph $G^N(V^N, E^N)$
**Initialization:** $G^N(V^N, E^N) \leftarrow G(V, E)$, $P_N \leftarrow \varphi$.
**Begin**
**While** ($\exists$ at least one *s-d* path in $G^N$)
  $p \leftarrow$ Minimum hop *s-d* path in $G^N$.
  $P_N \leftarrow P_N \cup \{p\}$
  $\forall_{\substack{vertex, v \in p \\ v \neq s,d \\ edge, e \in Adj\text{-}list(v)}} \quad G^N(V^N, E^N) \leftarrow G^N(V^N - \{v\}, E^N - \{e\})$
**end While**
**return** $P_N$
**End**

---

**Figure 2:** Algorithm to Determine the Set of Node-Disjoint *s-d* Paths in a Network Graph

---

**Input:** Graph $G(V, E)$, Source *s* and Destination *d*
**Output:** Set of Zone-Disjoint Paths $P_Z$
**Auxiliary Variables:** Graph $G^Z(V^Z, E^Z)$
**Initialization:** $G^Z(V^Z, E^Z) \leftarrow G(V, E)$, $P_Z \leftarrow \varphi$
**Begin**
**While** ($\exists$ at least one *s-d* path in $G^Z$)
  $p \leftarrow$ Minimum hop *s-d* path in $G^Z$
  $P_Z \leftarrow P_Z \cup \{p\}$
  $\forall_{\substack{vertex, u \in p, u \neq s,d \\ edge, e \in Adj\text{-}list(u)}} \quad G^Z(V^Z, E^Z) \leftarrow G^Z(V^Z - \{u\}, E^Z - \{e\})$
  $\forall_{\substack{vertex, u \in p, u \neq s,d \\ v \in Neighbor(u), v \neq s,d \\ edge, e' \in Adj\text{-}list(v)}} \quad G^Z(V^Z, E^Z) \leftarrow G^Z(V^Z - \{v\}, E^Z - \{e'\})$
**end While**
**return** $P_Z$
**End**

---

**Figure 3:** Algorithm to Determine the Set of Zone-Disjoint *s-d* Paths in a Network Graph

To determine the set $P_Z$ (refer Figure 3), we remove all the intermediate nodes (nodes other than the source *s* and destination *d*) that were part of the minimum hop *s-d* path *p* and also all their neighbour nodes from the original graph *G* to obtain the modified graph $G^Z(V^Z, E^Z)$. We determine the minimum hop *s-d* path in the modified graph $G^Z$, add it to the set $P_Z$ and remove the intermediate nodes that were part of this *s-d* path and all their neighbour nodes to obtain a new updated graph $G^Z(V^Z, E^Z)$. We then repeat this procedure until there exists no more *s-d* paths in the network. The set $P_Z$ is now said to contain the set of zone-disjoint *s-d* paths in the original network graph *G*. Note that when we remove a node *v* from a network graph, we also remove all the links associated with the node (i.e., links belonging to the adjacency list *Adj-list*(*v*)) where as when we remove a link from a graph, no change occurs in the vertex set of the graph.

The three algorithms could be implemented in a distributed fashion in ad hoc networks by flooding the route request (RREQ) message, letting the destination node to select and inform about the link-disjoint, node-disjoint and zone-disjoint routes to the source by using the route reply (RREP) packets. The source could then use these routes in the increasing order of hop count (i.e., use the least hop count route until it exists and then use the next highest hop count path as long as it exists and so on) or distribute the packets through several paths simultaneously with paths that have minimum hop count being used more.

## 4. SIMULATIONS

We ran our simulations in both square and rectangular network topologies of dimensions 1000m x 1000m and 2000m x 500m respectively. Both these network topologies have the same area. The average neighbourhood size is determined as follows: $N^* \pi R^2/A$, where *N* is the number of nodes in the network, *R* is the transmission range of a node and *A* is the network area. The transmission range per node used in all of our simulations is 250m. The simulations on both the square and rectangular network topologies were conducted for different values of the average node densities representing the neighbourhood size: 10 neighbours per node (50 nodes, low density), 20 neighbours per node (100 nodes, moderate density) and 30 neighbours per node (150 nodes, high density). By running the simulations in both square and rectangular network topologies, we also intend to study the impact of the variation in node distribution for a fixed value of average node density. Square topologies will have more uniform node distribution compared to rectangular topologies. We use the Random Waypoint mobility model [25], one of the most widely used models for simulating mobility in MANETs. According to this model, each node starts moving from an arbitrary location to a randomly selected destination with a randomly chosen speed in the range [$v_{min}$ .. $v_{max}$]. Once the destination is reached, the node stays there for a pause time and then continues to move to another randomly selected destination with a different speed. We use $v_{min} = 0$ and pause time of a node is also set to 0. The values of $v_{max}$ used are 10, 30 and 50 m/s representing low mobility, moderate mobility and high mobility levels respectively.

**Table 1:** Simulation Parameters

| Network Size (m x m) | 1000m x 1000m | 2000m x 500m |
|---|---|---|
| Number of Nodes | 50, 100 and 150 | |
| Transmission Range | 250m | |
| Simulation Time | 1000 seconds | |
| Number of Source-Destination (*s-d*) Pairs | 15 | |
| Topology Sampling Interval | 0.25 seconds | |
| Routing Strategies | Dijkstra algorithm [26] for minimum hop single path, Link-disjoint multi-path algorithm, Node-disjoint multi-path algorithm, Zone-disjoint multi-path algorithm | |
| Minimum Node Speed, $v_{min}$ | 0 m/s | |
| Maximum Node Speed, $v_{max}$ | 10 m/s (Low mobility scenario),<br>30 m/s (Moderate mobility scenario) and<br>50 m/s (High mobility scenario) | |

We obtain a centralized view of the network topology by generating mobility trace files for 1000 seconds under each of the above simulation conditions. We sample the network topology for every 0.25 seconds. Note that, two nodes *a* and *b* are assumed to have a bi-directional link at time *t,* if the Euclidean distance between them at time *t* (derived using the locations of the nodes from the mobility trace file) is less than or equal to the wireless transmission range of the nodes. Each data point in Figures 4 through 9 is an average computed over 10 mobility trace files and 15 *s-d* pairs from each of the mobility trace files. The starting time for each *s-d* session is uniformly distributed between 1 to 10 seconds. The simulation conditions are summarized in Table 1.

### 4.1 Determining the Sequence of Multi-path and Single Path Routes

We determine the sequence of link-disjoint, node-disjoint and zone-disjoint routes over the entire simulation time period as follows: When an *s-d* path is required at a given sampling time instant and there is none known, we run the appropriate multi-path algorithm to determine the set of disjoint routes for a given *s-d* pair. We assume the *s-d* routes in a disjoint multi-path set are used in the succeeding sampling time instants in the increasing order of the hop count. In other words, the *s-d* route with the next highest hop count is used as long as it exists and so on. We thus persist with the determined multi-path set of disjoint *s-d* routes as long as at least one path in the set exists. We repeat the above procedure till the end of the simulation time period. We also determine the sequence of single path *s-d* routes by running the minimum hop Dijkstra algorithm [26] on the network graph generated at the simulation time instant when an *s-d* route is used until it exists and the procedure is repeated over the duration of the network simulation session. The sequence of minimum-hop single path *s-d* routes is used as a benchmark to evaluate the relative increase in the time between multi-path route discoveries vis-à-vis single path discoveries and the corresponding increase in the average hop count for multi-path zone-disjoint, node-disjoint and link-disjoint routes.

### 4.2 Performance Metrics

We measure the following performance metrics:

- *Average Number of Paths per Multi-Path Set*: This is the number of disjoint paths (zone-disjoint or node-disjoint or link-disjoint, depending on the algorithm) determined during a multi-path route discovery, averaged over all the *s-d* sessions. In the case of single path routing, the number of paths determined per route discovery is 1.
- *Average Time between Successive Multi-Path/Single path Route Discoveries*: This is the time between two successive broadcast multi-path (or single path) discoveries, averaged across all the *s-d* sessions over the simulation time. As we opt for a route discovery only when all the paths in a multi-path set fails, this metric is a measure of the lifetime of the set of multi-paths and a larger value is preferred for a routing algorithm or protocol.
- *Average Hop Count per Multi-Path/Single path*: The average hop count for a given routing strategy is the time-averaged hop count of the individual paths that are used in a sequence over the entire simulation time period. For example, if the sequence of minimum hop paths used comprise of a 2-hop path for 2 seconds, then a 3-hop path for 3 seconds and then again a 2-hop path for 5 seconds, the time-averaged hop count of the single path routing strategy comprising the sequence of minimum hop paths over a 10-second simulation time period is (2*2+3*3+2*5)/10 = 2.3 seconds. Similarly, if the sequence of zone-disjoint paths determined comprise of a 2-hop path for 8 seconds, a 3-hop path for 3 seconds and a 4-hop path for 4 seconds, the time-averaged hop count of the zone-disjoint multi-path routing strategy over the 15-second simulation time period is (2*8+3*3+4*4)/15 = 2.7.

### 4.3 Average Number of Paths per Multi-path Set

In terms of absolute values of the number of multi-paths discovered per route discovery (refer Figures 4 and 5), at low network density, there are 4-6 link-disjoint/ node-disjoint paths in network with square topology and 2-4 link-disjoint/ node-disjoint paths in network with rectangular topology. In networks of moderate node density (refer Figure 4), there are 10-12 node-disjoint paths and 13-15 link-disjoint paths in square topology and close to 7-8 node-disjoint paths and 11-13 link disjoint paths in rectangular topology. In networks of high node density, there are close to 18 node-disjoint paths and 20-21 link-disjoint paths in square topology and close to 11-12 node-disjoint paths and 17-18 link-disjoint paths in rectangular topology. Thus, the number of link-disjoint/ node-disjoint paths discovered per route discovery increases significantly with increase in the network density. The multi-path route discovery approaches make use of the increase in the number of links and nodes as we increase the network density. There are more link-disjoint paths than node-disjoint paths in all the results, which makes sense.

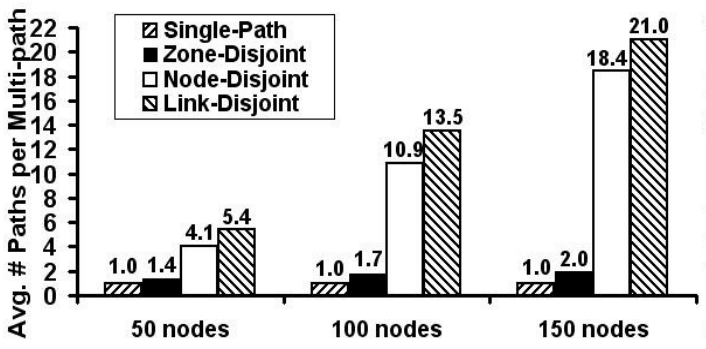

**Figure 4.1:** $v_{max}$ = 10 m/s

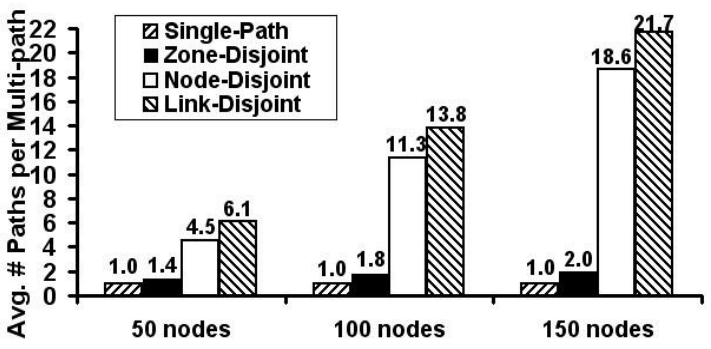

**Figure 4.2:** $v_{max}$ = 30 m/s

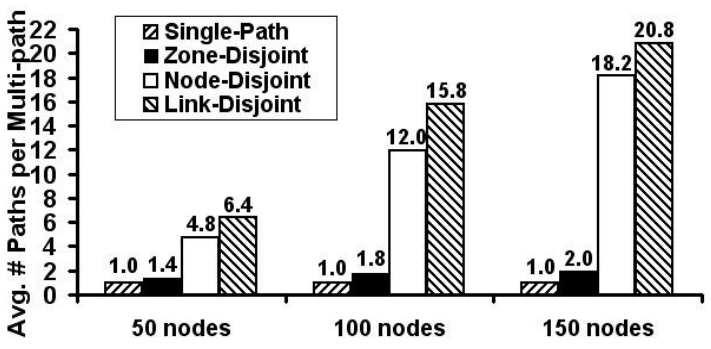

**Figure 4.3:** $v_{max}$ = 50 m/s

**Figure 4:** Average Number of Paths per Multi-path Set (1000m x 1000m Network)

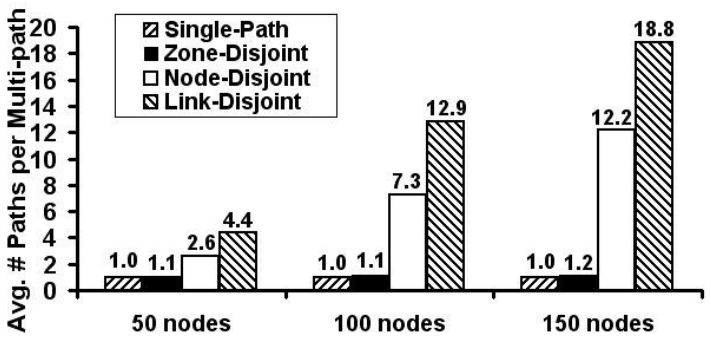

**Figure 5.1:** $v_{max}$ = 10 m/s

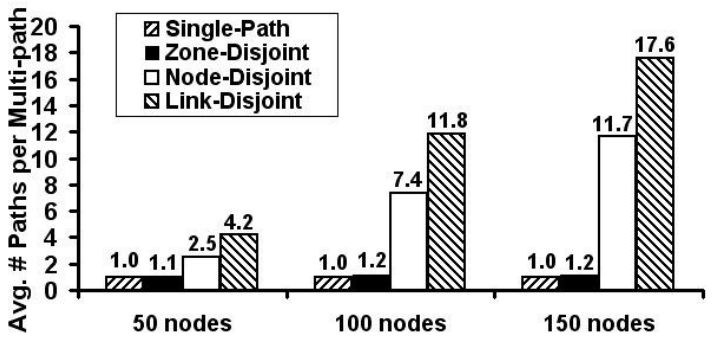

**Figure 5.2:** $v_{max}$ = 30 m/s

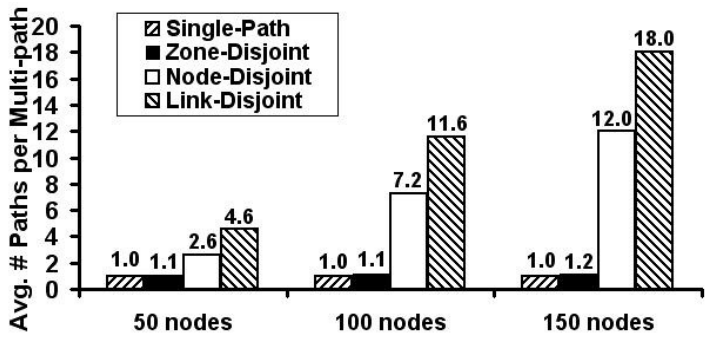

**Figure 5.3:** $v_{max}$ = 50 m/s

**Figure 5:** Average Number of Paths per Multi-path Set (2000m x 500m Network)

For a given network density, the difference in the number of link-disjoint paths and node-disjoint paths is more for a rectangular network topology than compared to those in a square topology. Also, for a given network density, the number of paths per route discovery is almost independent of node mobility for all the four types of routing strategies. For a fixed value of node mobility, the number of link-disjoint paths is more than the number of node-disjoint paths by a factor of 35% (square topology) to 70% (rectangular topology) in networks of low density, 20-35% (square topology) to 60-78% (rectangular topology) in networks of moderate density and 14-16% (square topology) to 50-54% (rectangular topology) in networks of high density. We also observe that for a given node density and level of node mobility, the number of multi-path routes in a square network topology is more than that obtained for a rectangular network topology. This can be attributed to the uneven distribution (distribution of more nodes in one direction compared to the other direction) in rectangular networks compared to square networks.

The number of paths per multi-path set for zone-disjoint routing has been observed to be significantly smaller than that observed for node-disjoint and link-disjoint routing. With zone-disjoint routing, when the intermediate nodes of the minimum hop path and also their neighbour nodes are removed from the network graph, the probability of an alternate path between the source and destination decreases significantly. In square networks, for a given node velocity, as we increase the network density from 50 nodes to 100 nodes, the average number of paths per multi-path set for zone-disjoint routing increases from 1.38 to 1.97 (43% increase); on the other hand, in rectangular networks, the average number of paths per multi-path set for zone-disjoint routing increases from 1.07 to 1.18 (10% increase). Thus, both in square and rectangular network topologies, there is no appreciable increase in the number of zone-disjoint paths, even with a three-fold increase in the network density. It is important to note that on the average, there can be at most two zone-disjoint paths and only one zone-disjoint path (i.e., nothing more than a single path) when we operate in square and rectangular networks with node density as large as 30 neighbours per node (i.e., in the 150 node network scenarios).

### 4.4 Average Time between Successive Multi-path and Single Path Route Discoveries

For a given network density and node mobility, the time between successive multi-path route discoveries incurred for each of the different multi-path routing strategies is low for a square network topology and is relatively high for a rectangular topology. As rectangular topologies become more one-dimensional, the hop count of the routes increases, thus resulting in more route breaks.

An interesting and significant observation is that the time between successive link-disjoint multi-path discoveries is at most 15% (square topology) – 25% (rectangular topology) larger than the time between successive node-disjoint multi-path discoveries. The difference in the time for two successive route discoveries between link-disjoint and node-disjoint routing decreases significantly with increase in the network density. In high density networks, there is no appreciable difference in the lifetime of the two multi-path routes, especially in a square network topology. The above observation illustrates that given a choice between the link-disjoint and node-disjoint strategies, it is worth to just opt for node-disjoint routes as they are have the highest aggregate bandwidth, provide the maximum possible fault-tolerance and also provide effective load balancing. The increase in stability comes with only a slight increase in the hop count (as observed in Section 4.5) compared to the minimum hop single path routing.

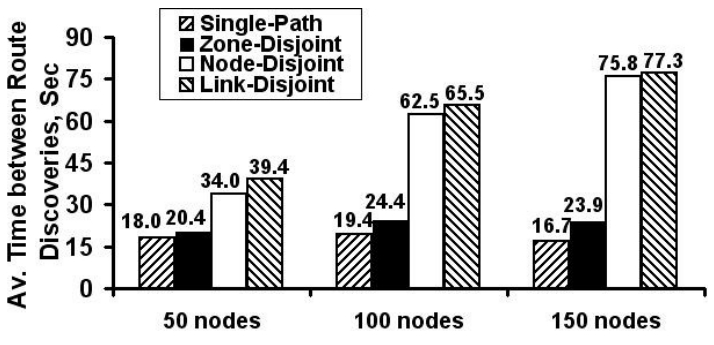

**Figure 6.1:** $v_{max}$ = 10 m/s

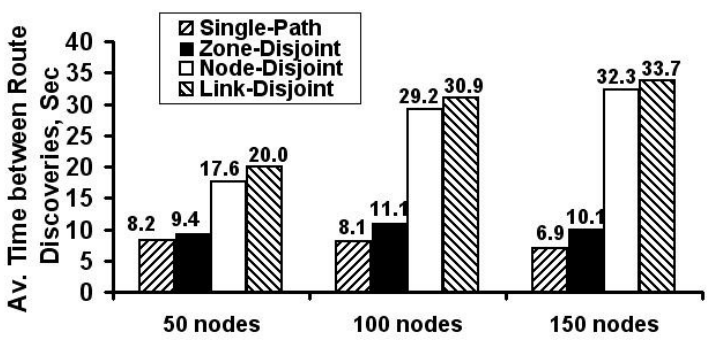

**Figure 6.2:** $v_{max}$ = 30 m/s

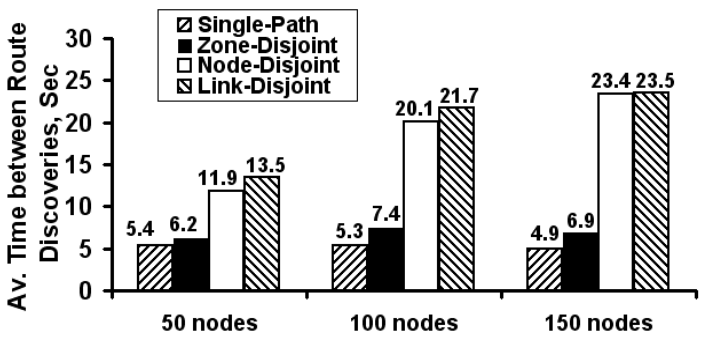

**Figure 6.3:** $v_{max}$ = 50 m/s

**Figure 6:** Average Time between Multi-path Route Discoveries (1000m x 1000m Network)

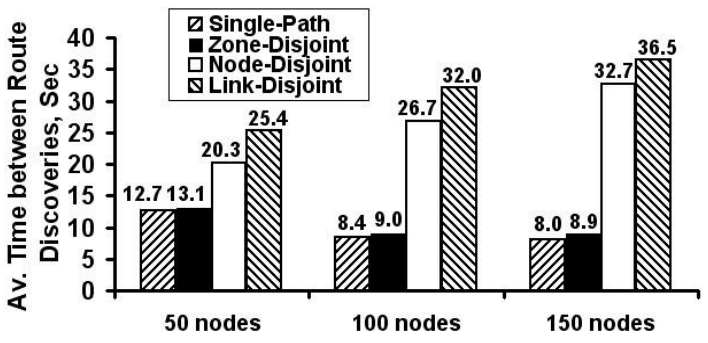

**Figure 7.1:** $v_{max}$ = 10 m/s

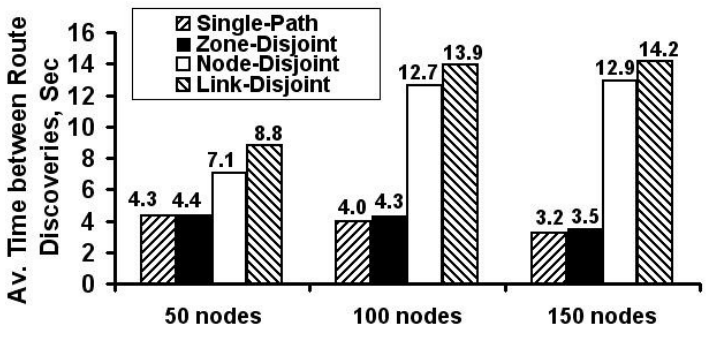

**Figure 7.2:** $v_{max}$ = 30 m/s

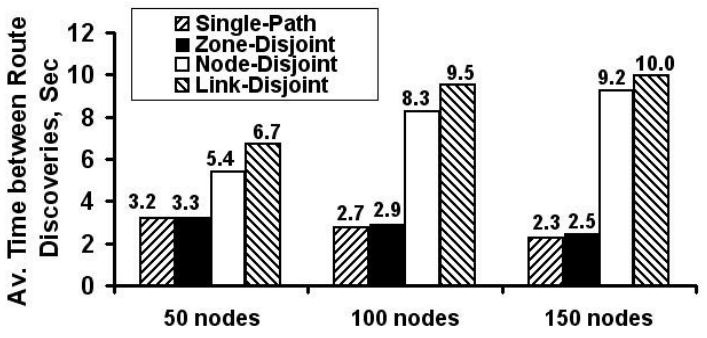

**Figure 7.3:** $v_{max}$ = 50 m/s

**Figure 7:** Average Time between Multi-path Route Discoveries (2000m x 500m Network)

For a given node mobility, the time between successive single path route discoveries decreases with increase in node density. This is due to the edge effect problem [27]. As, the number of nodes in a given neighbourhood increases, the minimum hop single path routing approach chooses the intermediate nodes that are as far away from each other so that the overall hop count is minimized. With increase in the network density, edge effect results in reduction of the hop count of single path routes by only 5-15%; the time between successive single path discoveries decreased by at most 15% and 35% in square and rectangular network topologies respectively.

For fixed node mobility, in low-density networks, the time between successive zone-disjoint, node-disjoint and link-disjoint route discoveries is 15%, 80-120% and 120-150% more than the time between single path route discoveries for the square network topology and is 3%, 60-70% and 100-110% more than the time between single path route discoveries for the rectangular network topology. In moderate-density networks, the time between successive zone-disjoint, node-disjoint and link-disjoint route discoveries is 25-40%, 220-275% and 240-300% more than the time between single path route discoveries for the square network topology and is 7-9%, 200-220% and 250-280% more than the time between single path route discoveries for the rectangular network topology. In high-density networks, the time between successive zone-disjoint, node-disjoint and link-disjoint route discoveries is 40-45%, 350-370% and 360-385% more than the time between single path route discoveries for the square network topology and is 10-11%, 300-310% and 340-350% more than the time between single path route discoveries for the rectangular network topology. In rectangular network topology, the time between successive route discoveries for single path routing and zone-disjoint routing is almost the same with the latter being at most 11% more than the former. Thus, we cannot significantly reduce the route discovery control overhead with zone-disjoint multi-path routing.

Even though we observe a direct correlation between the number of paths per multi-path set and the time between successive multi-path route discoveries, for zone-disjoint, node-disjoint and link-disjoint routing, the increase in the number of paths per multi-path set with increase in node density does not yield a corresponding proportional increase in the time between successive multi-path route discoveries. For example, in square network topologies, even though the number of zone-disjoint paths per multi-path set increases from 1.38 to 1.97 with increase in node density from 10 to 30 neighbours per node, the time between successive zone-disjoint multi-path route discoveries can be at most 20% larger. For node-disjoint path routing, as we increase node density from 10 neighbours per node to 30 neighbours per node, even though the absolute value for the number of paths per multi-path set increases from 4.4 to 18.4, the time between successive multi-path route discoveries increases only by at most 120%. Similar observations can be made for link-disjoint routing.

For a rectangular network topology, the increase in the time between successive multi-path route discoveries with increase in node density from 10 to 30 neighbours per node is relatively low compared to that incurred with a square network topology. This can be also attributed to the relatively unstable nature of the minimum-hop routes in rectangular network topologies compared to square network topologies. The minimum-hop routes in rectangular network topologies have a larger hop count (explained more in Section 4.5) compared to those incurred with square network topologies. Each of the links in a minimum hop path has almost the same probability of failure in both square and rectangular network topologies [28]. As a result, since there are more hops, the probability of failure of a minimum hop path is more in rectangular network topologies compared to square network topologies. The impact of the topology shape on the stability of the routes is also vindicated by the relatively rapid decrease in the lifetime per multi-path set in rectangular network topologies with increase in the level of node mobility compared to that incurred in square network topologies.

### 4.5 Average Hop Count per Multi-path and Single Path

For networks of square topology, the average hop count of the sequence of link-disjoint routes is almost the same as the average hop count for the sequence of node-disjoint routes. For networks of rectangular topology, the average hop count of link-disjoint routes is only 5-10% more than that of node-disjoint routes. This is a significant observation as the node-disjoint routes have smaller hop counts and hence could yield lower end-to-end delay per packet. For any level of node mobility and node density, the average hop count per zone-disjoint multi-path set can be at most 10% (for square network topology) and 3% (for rectangular network topology) more than that of the minimum hop count obtained via single path routing. Thus, there is relatively insignificant difference in the hop count incurred by the zone-disjoint and the single path routing strategies. For a given network density, the hop count of the routes does not change significantly with node mobility. In networks of low density, the average hop count of the link-disjoint/ node-disjoint routes is only at most 5% more than that of the minimum hop single path routes. In networks of moderate and high density, the average hop count of the link-disjoint/ node-disjoint routes is still only 10-25% more than that of the minimum hop single path routes. Thus, with increase in network density, even though the number of link-disjoint/ node-disjoint routes discovered per route discovery increases significantly, the average hop count of these routes does not significantly increase when compared to those incurred in minimum hop single path routing.

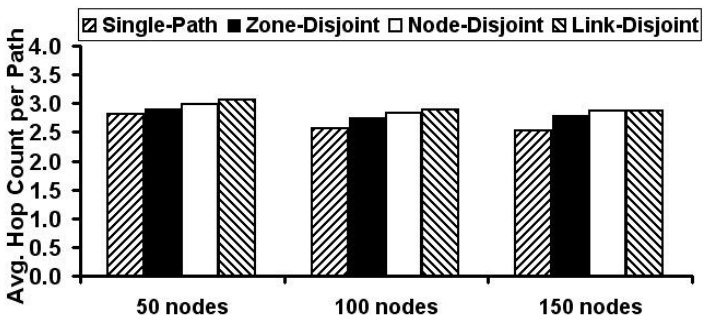

**Figure 8.1:** $v_{max}$ = 10 m/s

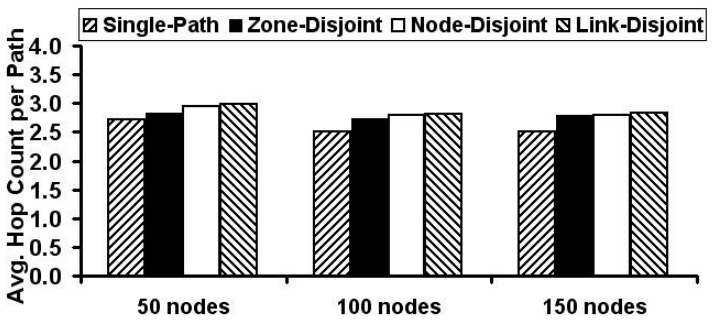

**Figure 8.2:** $v_{max}$ = 30 m/s

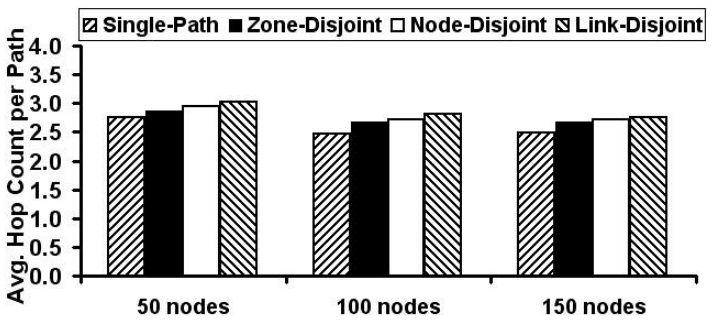

**Figure 8.3:** $v_{max}$ = 50 m/s

**Figure 8:** Average Hop Count per Multi-Path/ Single Path (1000m x 1000m Network)

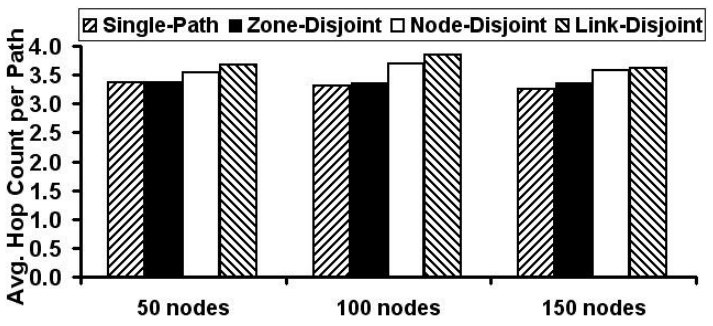

**Figure 9.1:** $v_{max}$ = 10 m/s

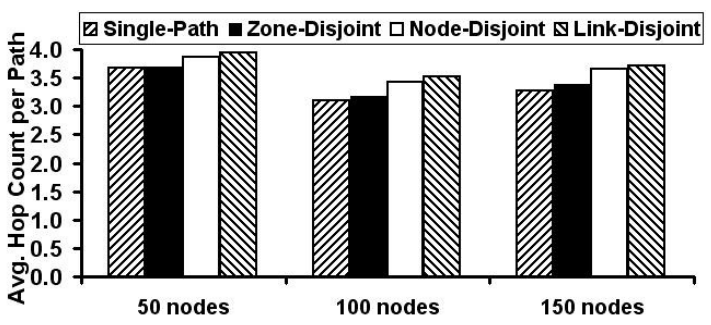

**Figure 9.2:** $v_{max}$ = 30 m/s

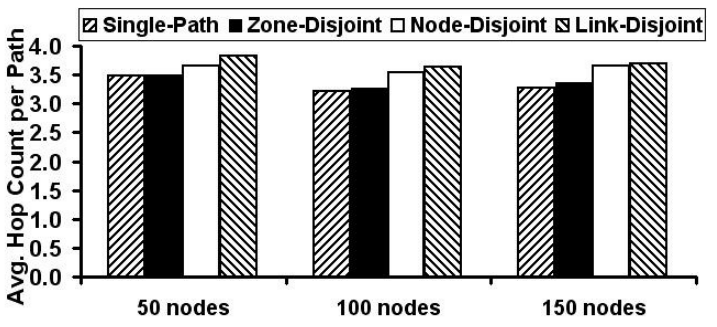

**Figure 9.3:** $v_{max}$ = 50 m/s

**Figure 9:** Average Hop Count per Multi-Path/ Single Path (2000m x 500m Network)

In terms of the absolute numbers, for a given level of node mobility, routing strategy and network topology, the average hop count incurred with the multi-path routes as well as the single path routes decreases with increase in node density. The decrease is more predominant (by a factor of at most 10%) with the single path routing strategy compared to the three multi-path routing strategies (only by a factor of at most 5%). This can be attributed to the fact that the constituent routes of the zone-disjoint, node-disjoint and link-disjoint multi-paths may not be minimum hop routes. Also, both the number of paths per multi-path set and the time between successive multi-path route discoveries increase with increase in node density. The above observation is especially more relevant for node and link-disjoint multi-path routing.

We also observe that for a given level of node mobility and node density, the average hop count per routing strategy in a rectangular network topology can be 40% - 50% more than that incurred in a square network topology. This can be attributed to the fact that in the rectangular network topologies the nodes are more predominantly distributed in one-dimension (actually in the longer of the two dimensions) and this contributes to the relatively larger hop count compared to the square network topologies where nodes are more uniformly distributed [28]. The relatively larger hop count contributes to the unstable nature of the minimum hop routes in rectangular network topologies compared to those discovered in square network topologies.

## 5. CONCLUSIONS AND FUTURE WORK

We analyzed the performance of link, node and zone-disjoint multi-path routing algorithms vis-à-vis single path minimum hop routing with respect to performance metrics such as the time between successive multi-path route discoveries, the hop count per multi-path set and the multi-path set size. A significant observation is that the link-disjoint multi-paths are only 15-30% more stable compared to node-disjoint multi-paths with often negligible difference in the average hop count. Simulation results indicate that with an average neighbourhood size of 10, the time between successive minimum-hop single path route discoveries is around 50-75% of the time between successive node-disjoint and link-disjoint multi-path route discoveries; whereas, with an average neighbourhood size of 30, the time between successive minimum-hop single path route discoveries is only 20-25% of the time between successive link-disjoint/ node-disjoint multi-path route discoveries. At the same time, the average hop count in a sequence of node-disjoint/ link-disjoint multi-paths is only 10-20% more than that of a sequence of minimum-hop single path routes.

Based on the simulation results obtained in this paper, one could conclude that, on average, the number of zone-disjoint paths can be as large as 2 and the time between successive zone-disjoint multi-path discoveries can be at most 42% (for square topologies) and 10% (for rectangular topologies) more than that incurred with single path routing. On the other hand, the time between successive node-disjoint and link-disjoint multi-path route discoveries can be significantly larger than that incurred with zone-disjoint routing. The corresponding increase in the average hop count per node-disjoint multi-path set and link-disjoint multi-path set is only 13% more than that of the minimum hop single paths. Also, the worst-case difference in the average hop count per zone-disjoint multi-path set compared to node-disjoint multi-path set and link-disjoint multi-path sets is within 5% and this is relatively insignificant compared to the significant reduction in the route discovery overhead that can be potentially brought about through node-disjoint routing.

As future work, we would develop distributed routing protocols based on our zone-disjoint, node-disjoint and link-disjoint routing algorithms and compare the three routing protocols with respect to metrics such as throughput and end-to-end delay. We will study the benefits and drawbacks associated in simultaneously routing through at most two zone-disjoint paths vis-à-vis routing through multiple node-disjoint paths and link-disjoint paths. Future work would also involve analyzing the energy consumption aspect of multi-path routing and studying the effect on node lifetime.